\documentclass[jkps,fleqn,showkeys]{revtex4}
\usepackage{amssymb}
\usepackage{amsmath}
\usepackage{bm}
\usepackage{graphicx}
\usepackage{tikz}
\usetikzlibrary{arrows, positioning, fit, backgrounds, shapes}
\usepackage{placeins}

\begin{document}
\setcounter{page}{0}
\title[]{Inferring Coupling Strength from the Kuramoto Order Parameter}
\author{Gug Young \surname{Kim}}
\affiliation{Department of Applied Physics, Hanyang University, Ansan, 15588, Republic of Korea}
\author{Hoseok \surname{Sul}}
\affiliation{Department of Marine Science and Convergence Technology, Hanyang University, Ansan, 15588, Republic of Korea}
\author{Jee Woong \surname{Choi}}
\email{choijw@hanyang.ac.kr}
\affiliation{School of Defense Intelligence and Information Convergence Engineering, Hanyang University, Ansan, 15588, Republic of Korea}
\author{Seung-Woo \surname{Son}}
\email{sonswoo@hanyang.ac.kr}
\affiliation{Department of Applied Physics, Hanyang University, Ansan, 15588, Republic of Korea}

\date[]{Received \today}

\begin{abstract}
Accurately estimating the coupling strength in oscillator networks from macroscopic observations is essential for predicting synchronization transitions. We consider the inverse problem of reconstructing the unknown coupling strength $K$ in the globally coupled Kuramoto model from scalar observations of the macroscopic order parameter $R(t)$, assuming that the natural frequencies and the initial phase configuration are known. After initialization, individual phase trajectories are treated as hidden, and only the scalar order parameter is observed. We employ an extended Kalman filter with an augmented state representation that recursively estimates the coupling strength from observations of $R(t)$. By exploiting the mean-field structure of the globally coupled Kuramoto model, the covariance prediction step can be computed efficiently, substantially reducing the computational cost. Numerical simulations demonstrate that the proposed estimator accurately reconstructs the coupling strength and remains stable even when $R(t)$ is small and strongly fluctuating.
\end{abstract}


\maketitle

\section{Introduction}
Synchronization is a ubiquitous collective phenomenon observed in a wide range of natural and engineered systems, including biological rhythms, chemical oscillators, power grids, and coupled nonlinear oscillators. Since the pioneering studies of Winfree and Kuramoto, the Kuramoto model has served as a paradigmatic framework for describing how microscopic phase interactions give rise to macroscopic synchronization transitions~\cite{Winfree1967,Kuramoto1975,Strogatz2000,Acebron2005,Rodrigues2016}. In the standard forward problem, the microscopic parameters of the oscillator population, such as the natural frequencies and coupling strength, are given, and the resulting macroscopic order parameter is predicted. In contrast, the inverse problem considered here asks whether an unknown microscopic interaction parameter can be inferred from limited macroscopic observations.

Estimating the coupling strength of a synchronized oscillator population is particularly important because it determines the onset and stability of collective coherence. 
Moreover, in oscillator systems with effective inertia, the coupling strength can also delimit oscillatory regimes of global synchrony induced by secondary synchronized clusters~\cite{Kim2026CSF}. Existing inference approaches using microscopic time-series data, such as the phases or signals of individual oscillators, can reconstruct phase dynamics, coupling functions, or network connectivity~\cite{Timme2007,Kralemann2011,Tirabassi2015}. Such approaches are powerful but require access to sufficiently resolved individual oscillator trajectories.

We consider a measurement protocol in which microscopic information is available at initialization, while subsequent inference uses only a coarse-grained collective signal. In populations of coupled electrochemical oscillators, the Kuramoto order parameter has been extracted from global measurements rather than from the complete time series of all individual oscillators~\cite{Zhai2005}. Macroscopic descriptions are also useful for biological oscillator populations such as circadian systems~\cite{Hannay2018,Schmal2018}. While individual clock-cell trajectories can in principle be observed at the single-cell level~\cite{Abel2016}, such approaches are often costly and may introduce non-negligible measurement noise into the system. Another related example arises in stimulus-evoked neural oscillations~\cite{Telenczuk2010}. An external stimulus can induce a simultaneous phase reset across a population of microscopic neuronal sources, thereby defining a known initial phase relationship at stimulus onset, whereas only the macroscopic EEG/MEG signal can be recorded noninvasively thereafter, since individual neuronal trajectories remain experimentally inaccessible in vivo. These examples motivate the macroscopic-observation setting considered in this study: after initialization, the microscopic phases are treated as hidden variables, and only the scalar order parameter is supplied to the estimator. It is worth noting that this is a modeling choice regarding the information made available to the estimator, rather than a claim about the experimental accessibility of individual trajectories.

Some macroscopic approaches to coupling-strength estimation rely on the relaxation dynamics of the order parameter near a synchronized state. In such cases, the coupling strength $K$ can be inferred from the characteristic relaxation time following a perturbation~\cite{Son2008}. Macroscopic inference is also possible in nonsynchronized regimes~\cite{Kato2025,Yamaguchi2024}. For the Kuramoto model with a unimodal frequency distribution, the thermodynamic-limit stationary order parameter vanishes below the critical coupling strength $K_c$~\cite{Acebron2005,Dorfler2011}. In finite-size systems, the order parameter does not vanish identically but exhibits sample-dependent temporal fluctuations~\cite{Hong2015}. These fluctuations are usually regarded as finite-size noise, yet they may still encode some information about the underlying coupling strength when combined with a dynamical model.

Recent studies have also explored parameter estimation from macroscopic quantities and data-assimilation approaches for coupled oscillator systems~\cite{Kato2025, SmithGottwald2025}. Kato et al.~\cite{Kato2025} estimate coupling strength and the width of a Lorentzian frequency distribution from macroscopic trajectories using Bayesian inference, whereas Smith and Gottwald~\cite{SmithGottwald2025} estimate hidden phases and individual natural frequencies from partial microscopic observations using an ensemble Kalman filter. Yamaguchi and Terada~\cite{Yamaguchi2024} reconstruct phase dynamics from macroscopic responses to weak external forcing, while Choi and Yamaguchi~\cite{Choi2026} combine macroscopic responses with an individual oscillator's amplitude response. Computationally, these approaches involve likelihood evaluations, ensemble propagation, or response inversion. 

The present study develops a recursive extended Kalman filter (EKF) based estimator that uses only the scalar order parameter $R(t)$ after initialization, without external forcing and with known initial phases and individual natural frequencies. Here we propose an EKF framework~\cite{Kalman1960,Jazwinski1970} for estimating the unknown coupling strength $K$ from the scalar macroscopic observation $R(t)$. We assume that the natural frequencies and the initial phase configuration are known at the initialization stage, while the true coupling strength is unknown. After initialization, the microscopic phase trajectories are not observed, only the time series of the scalar order parameter is received. By recursively propagating the microscopic state estimate and correcting it using $R(t)$, the EKF extracts hidden dynamical correlations between finite-size order-parameter fluctuations and the coupling strength.

A direct EKF implementation for an $N$-oscillator system is computationally expensive because the covariance prediction step involves dense matrix multiplications with cubic complexity. To overcome this limitation, we exploit the mean-field structure of the globally coupled Kuramoto model. We show that the phase-interaction block of the Jacobian can be exactly decomposed into a diagonal component and a rank-2 perturbation. This structure allows the covariance prediction to be evaluated using matrix-vector products and vector outer products, reducing the leading computational cost from $\mathcal{O}(N^3)$ to $\mathcal{O}(N^2)$. Numerical simulations demonstrate that the proposed method can estimate $K$ using only macroscopic observations and remains bounded even in the subcritical regime, where the order parameter is dominated by finite-size fluctuations.

In this study, we restrict our analysis to the idealized noiseless-observation setting, in which the scalar observation $z_k$ coincides exactly with the simulated order parameter $R(t)$. This choice allows us to isolate and characterize the intrinsic information content of finite-size order-parameter fluctuations for coupling-strength inference, independent of measurement-noise effects. A systematic study of robustness to observation noise is left for future work.

The remainder of this paper is organized as follows. Section II defines the globally coupled Kuramoto model and the macroscopic-observation setting, and formulates the EKF state-space model using an augmented state vector. Section III exploits the mean-field structure of the model to decompose the phase-interaction block of the Jacobian exactly into a diagonal component and a rank-2 perturbation. Section IV presents numerical results, including representative estimation trajectories, ensemble-level accuracy and its relation to finite-size order-parameter fluctuations, and a computational benchmark demonstrating the efficiency of the rank-2 update. Finally, Section V summarizes the findings and discusses the limitations of the present framework together with directions for future work.

\section{Problem Setup and State-Space Formulation}

\begin{figure*}[t!]
    \centering
    \includegraphics[width=1\linewidth]{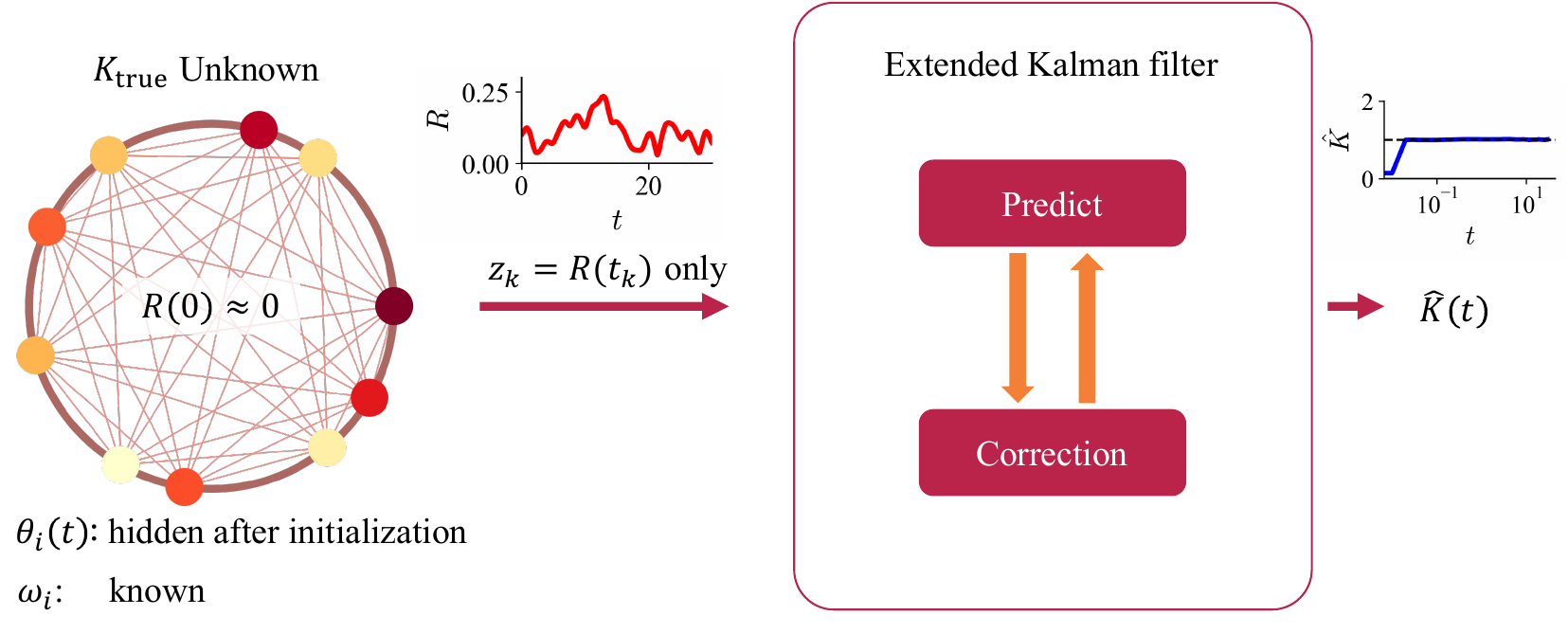}
    \caption{
    Schematic illustration of the EKF-based coupling-strength inference.
    The initial phases and natural frequencies are assumed to be known, while the coupling strength
    $K_{\mathrm{true}}$ is unknown.
    After initialization, only the scalar order parameter $z_k=R(t_k)$ is observed.
    The EKF internally propagates the microscopic phase estimates and recursively updates the
    coupling-strength estimate $\hat{K}(t)$.
    }
    \label{fig:concept}
\end{figure*}

We consider a population of $N$ globally coupled phase oscillators described by the Kuramoto model,
\begin{equation}
    \dot{\theta}_i
    =
    \omega_i
    +
    \frac{K_{\mathrm{true}}}{N}
    \sum_{j=1}^{N}
    \sin(\theta_j-\theta_i),
    \qquad i=1,\ldots,N,
    \label{eq:kuramoto}
\end{equation}
where $\theta_i(t)$ and $\omega_i$ denote the phase and natural frequency of the $i$-th oscillator, respectively. The parameter $K_{\mathrm{true}}$ is the true coupling strength to be inferred.
The complex Kuramoto order parameter quantifies the collective synchronization state,
\begin{equation}
    Z(t)
    =
    R(t)e^{\mathrm{i}\Psi(t)}
    =
    \frac{1}{N}
    \sum_{j=1}^{N}
    e^{\mathrm{i}\theta_j(t)},
    \label{eq:order_parameter}
\end{equation}
where $R(t)\in[0,1]$ measures the degree of phase coherence and $\Psi(t)$ is the mean phase. Equivalently, by defining
\begin{equation}
    X(t)
    =
    \frac{1}{N}
    \sum_{j=1}^{N}\cos\theta_j(t),
    \qquad
    Y(t)
    =
    \frac{1}{N}
    \sum_{j=1}^{N}\sin\theta_j(t),
    \label{eq:xy_order_parameter}
\end{equation}
the scalar order parameter is written as $R(t)=\sqrt{X^2(t)+Y^2(t)}$.

The inverse problem considered here is to estimate $K_{\mathrm{true}}$ from the scalar macroscopic time series $R(t)$. We assume that the natural frequencies and the microscopic initial phase configuration are known at the initialization stage, $\hat{\theta}_{i,0}=\theta_i(0)$, $\hat{\omega}_i=\omega_i$, whereas the coupling strength is unknown and the estimator is initialized with an initial guess, $\hat{K}_0=K_{\mathrm{init}}=0$, chosen independently from $K_{\mathrm{true}}$. After initialization, the microscopic phase trajectories $\theta_i(t)$ are not observed. The only information supplied to the estimator is the scalar order parameter sampled at discrete times, $z_k = R(t_k)$. Thus, the estimator must infer the hidden coupling strength by propagating the microscopic phase estimates internally and comparing the predicted macroscopic order parameter with the observed scalar signal.

To formulate the estimation problem, we introduce the augmented state vector
\begin{equation}
    x_k
    =
    \left[
    \theta_{1,k},
    \theta_{2,k},
    \ldots,
    \theta_{N,k},
    K_k
    \right]^T .
    \label{eq:augmented_state}
\end{equation}
In the EKF prediction model, the phase dynamics are discretized using the Euler method with time step $\Delta t$,
\begin{equation}
    \theta_{i,k+1}
    =
    \theta_{i,k}
    +
    \Delta t
    \left[
    \omega_i
    +
    K_k
    \left(
    Y_k\cos\theta_{i,k}
    -
    X_k\sin\theta_{i,k}
    \right)
    \right],
    \label{eq:discrete_theta}
\end{equation}
where
\begin{equation}
    X_k
    =
    \frac{1}{N}
    \sum_{j=1}^{N}\cos\theta_{j,k},
    \qquad
    Y_k
    =
    \frac{1}{N}
    \sum_{j=1}^{N}\sin\theta_{j,k}.
    \label{eq:discrete_xy}
\end{equation}
The coupling strength is treated as a static but unknown parameter and is modeled as a random walk,
\begin{equation}
    K_{k+1}=K_k+\eta_k,
    \qquad
    \eta_k\sim\mathcal{N}(0,Q_K).
    \label{eq:K_random_walk}
\end{equation}
Here, $\eta_k$ is zero-mean Gaussian process noise introduced in the EKF prediction model for the coupling strength, and $Q_K$ is its variance. We use $Q_K=10^{-10}$ in the numerical experiments.
The corresponding observation model is
\begin{equation}
    z_k
    =
    R_k+\epsilon_k,
    \qquad
    {\rm where}~~~R_k=\sqrt{X_k^2+Y_k^2},
    ~~\epsilon_k\sim\mathcal{N}(0,R_c). 
    \label{eq:observation_model}
\end{equation}
Here, $\epsilon_k$ is zero-mean Gaussian observation noise, and $R_c$ is its variance.

In the synthetic-data experiments below, the observation noise was set to zero, so that $z_k$ corresponds directly to the sampled order parameter $R(t_k)$. Because the goal of this study is to determine whether $K_{\rm true}$ can be recovered from the intrinsic finite-size fluctuations of $R(t)$ under ideal measurement conditions, we set the observation noise to zero throughout the numerical experiments ($R_c=0$). The general noisy observation model of Eq. (8), together with the corresponding EKF update equations in Appendix A, is retained in full generality and provides a direct basis for extending the present framework to noisy measurements in future work.
In the numerical experiments, the true Kuramoto dynamics used to generate $R(t_k)$ are integrated using a fourth-order Runge--Kutta method. In contrast, the EKF uses the Euler-discretized state-space model in Eq.~\eqref{eq:discrete_theta} for prediction. The detailed EKF prediction and measurement-update equations are summarized in Appendix~A.

Figure~\ref{fig:concept} provides a schematic summary of the estimation setup described above. The initial oscillator phases and natural frequencies are assumed to be known and are supplied to the EKF at initialization, whereas the true coupling strength $K_{\rm true}$ remains unknown. After this initialization stage, the filter no longer has access to the individual phase trajectories; instead, it receives only the scalar order-parameter observations $z_k=R(t_k)$ at each time step. Internally, the EKF propagates its estimate of the microscopic phase configuration forward using the state-space model and uses the mismatch between the predicted and observed order parameter to recursively correct both the phase estimates and the coupling-strength estimate $\hat{K}(t)$, as illustrated by the feedback loop in the figure.

\section{Low-Rank EKF for Globally Coupled Oscillators}

A direct implementation of the EKF for the augmented state
$\mathbf{x}_k=[\theta_{1,k},\ldots,\theta_{N,k},K_k]^T$
requires the prediction of the covariance matrix in Appendix~A,
\begin{equation}
    P_{k|k-1}
    =
    F_{k-1}P_{k-1|k-1}F_{k-1}^{T}+Q.
\end{equation}
Here, $P_{k|k-1}$ is the predicted error covariance of the augmented state before the measurement update using $z_k$, and $P_{k-1|k-1}$ is the posterior error covariance after the measurement update using $z_{k-1}$. The matrix $F_{k-1}$ is the Jacobian of the discrete-time state-transition model evaluated at the previous posterior state estimate, and $Q$ is the process-noise covariance matrix. These definitions follow standard EKF notation~\cite{Kalman1960,Jazwinski1970}.
Since the phase variables are globally coupled, the Jacobian matrix $F_k$ is generally dense.
A naive evaluation of this covariance prediction therefore requires dense matrix-matrix multiplications, leading to a computational cost of $\mathcal{O}(N^3)$.
This cubic scaling becomes a serious bottleneck when the number of oscillators is large.

The mean-field structure of the globally coupled Kuramoto model allows this cost to be reduced.
The Jacobian matrix of the augmented state can be partitioned as
\begin{equation}
    F_k
    =
    \begin{bmatrix}
    A_k & \mathbf{b}_k \\
    \mathbf{0}^T & 1
    \end{bmatrix},
\end{equation}
where $A_k\in\mathbb{R}^{N\times N}$ is the phase-interaction block and
$\mathbf{b}_k\in\mathbb{R}^{N}$ is the sensitivity of the phase update with respect to $K_k$.
The key observation is that $A_k$ is not an arbitrary dense matrix.
By using the mean-field variables $X_k$ and $Y_k$, it can be decomposed exactly into a diagonal component and a rank-2 perturbation,
\begin{equation}
    A_k
    =
    D_k
    +
    \alpha
    \left(
    \mathbf{c}\mathbf{c}^{T}
    +
    \mathbf{s}\mathbf{s}^{T}
    \right),
    \qquad
    \alpha=\frac{\Delta t K_k}{N},
    \label{eq:rank2_main}
\end{equation}
where $\mathbf{c} = (\cos\theta_{1,k},\ldots,\cos\theta_{N,k})^T$ and $\mathbf{s} = (\sin\theta_{1,k},\ldots,\sin\theta_{N,k})^T$.
The diagonal matrix $D_k$ has entries $(D_k)_{ii}=1-\Delta t K_k(X_k\cos\theta_{i,k}+Y_k\sin\theta_{i,k})$, as given in Eq.~\eqref{eq:D_diagonal} in Appendix~\ref{app:rank2}.

This rank-2 structure makes it unnecessary to form and multiply the full dense Jacobian.
Terms such as $A_{k-1}P_{\theta\theta,k-1}A_{k-1}^{T}$ can be evaluated using matrix-vector products, scalar contractions, and vector outer products, where $P_{\theta\theta,k-1}$ is the phase-phase block of the posterior covariance at step $k-1$.
As a result, the leading computational cost of the covariance prediction is reduced from $\mathcal{O}(N^3)$ to $\mathcal{O}(N^2)$.
The detailed derivation of the Jacobian elements and the rank-2 covariance prediction is provided in Appendix~B.

\section{Numerical Results}


\begin{figure*}[b!]
    \centering
    \includegraphics[width=0.8\linewidth]{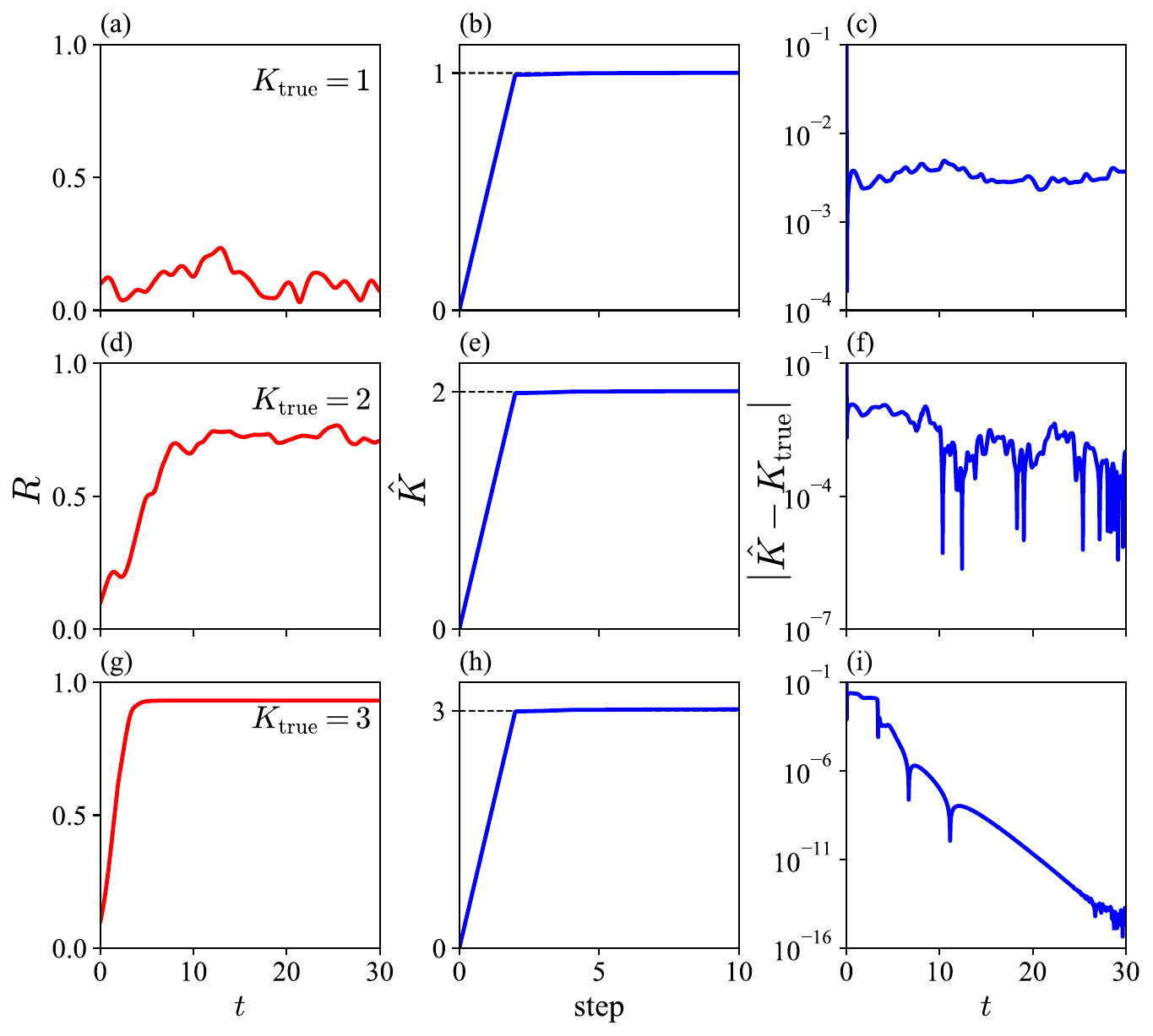}
    \caption{
    Representative tracking dynamics for $K_{\mathrm{true}}=1$, $2$, and $3$ with $N=200$.
    Panels (a), (d), and (g) show the observed order parameter $R(t)$.
    Panels (b), (e), and (h) show the early evolution of the coupling-strength estimate $\hat K(t)$ during the first ten EKF update steps.
    Panels (c), (f), and (i) show the absolute estimation error
    $|\hat K(t)-K_{\mathrm{true}}|$ on a logarithmic scale.
    The horizontal dashed lines in the middle column indicate the corresponding true coupling strengths $K_{\mathrm{true}}$.
    The estimator remains bounded in the fluctuating regime at $K_{\mathrm{true}}=1$, while it converges more accurately for larger coupling strengths.}
    \label{fig:tracking}
\end{figure*}

We numerically evaluated the proposed EKF-based estimator using synthetic data generated from the Kuramoto model.
The true Kuramoto dynamics were integrated using a fourth-order Runge--Kutta method with time step $\Delta t=0.01$.
The natural frequencies were sampled from a standard Gaussian distribution, $\omega_i\sim\mathcal{N}(0,1)$,
and the initial phases were sampled independently from a uniform distribution, $\theta_i(0)\sim U[0,2\pi)$.
This choice gives a disordered initial condition with $R(0)\approx0$.
Unless otherwise stated, the representative tracking and ensemble-estimation experiments were performed with $N=200$ oscillators.
For the standard Gaussian distribution, $g(\omega) = \exp\left(-{\omega^2}/{2}\right) /{\sqrt{2\pi}}$,
the critical coupling in the thermodynamic limit is $K_c = {2}/{\pi g(0)} = \sqrt{{8}/{\pi}} \simeq 1.596$.

In the estimation procedure, the natural frequencies and the initial phase configuration were supplied to the EKF, whereas the coupling strength was initialized with the initial guess $K_{\mathrm{init}}=0$.
After initialization, only the scalar order-parameter time series $R(t_k)$ was used for measurement updates.
The EKF prediction step used the Euler-discretized state-space model in Eq.~\eqref{eq:discrete_theta}. Full details of the covariance initialization, process-noise settings, and the projection bound on $\hat{K}$ are given in Appendix C.
Unless otherwise stated, the simulations were performed using the no-lock version of the EKF, in which $\hat{K}$ is updated continuously throughout the entire time interval without applying a freeze or locking condition.


Figure~\ref{fig:tracking} shows representative estimation trajectories for
$K_{\mathrm{true}}=1,2,$ and $3$, corresponding respectively to subcritical, intermediate, and strongly synchronized regimes relative to $K_c\simeq1.596$.
In all cases, the EKF was initialized with the correct microscopic initial phases and natural frequencies but with an incorrect initial estimate of $K$.
After initialization, only the scalar time series $R(t_k)$ was supplied to the filter.

For $K_{\mathrm{true}}=1$, the order parameter remains small and fluctuating, but the estimated coupling strength remains bounded.
For larger coupling strengths, the macroscopic signal becomes more coherent and the residual estimation error decreases more rapidly.
These results indicate that the EKF can extract information about $K_{\mathrm{true}}$ not only from strongly synchronized trajectories but also from finite-size fluctuations in the weakly synchronized or subcritical regime. A representative example of the underlying microscopic phase-tracking accuracy is shown in Appendix D, where the estimated and true phase trajectories remain closely aligned even though only the scalar order parameter $R(t)$ is observed after initialization.

\begin{figure*}[t!]
    \centering
    \includegraphics[width=1\linewidth]{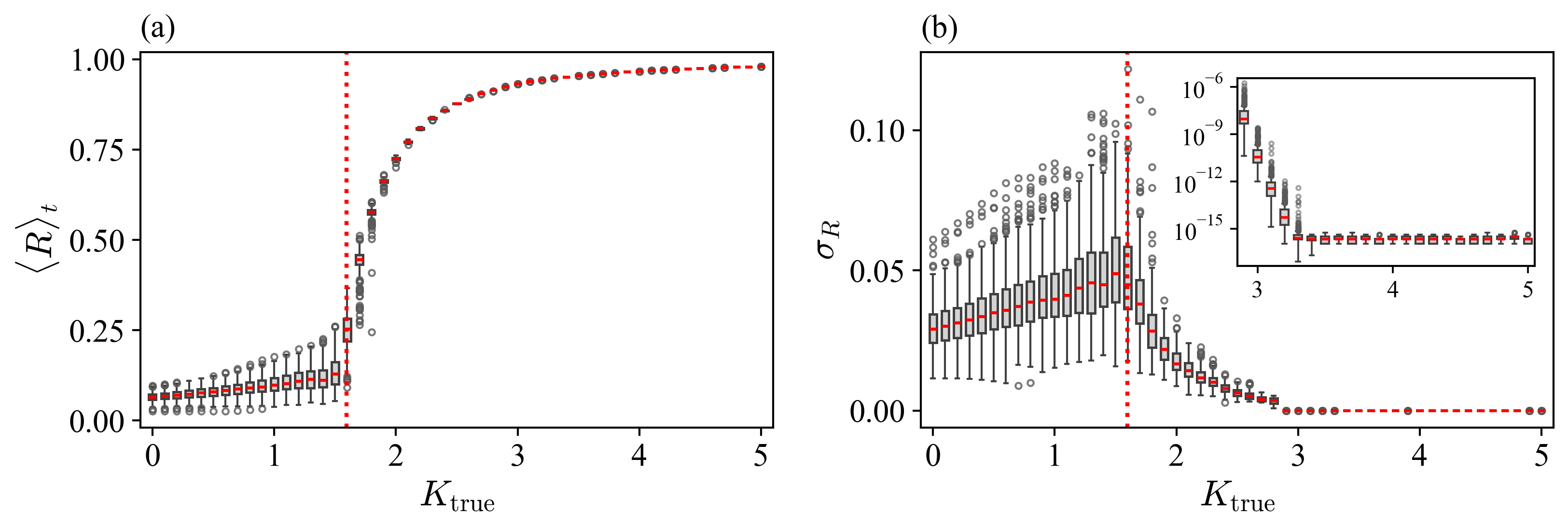}
    \caption{
    Finite-size macroscopic behavior of a globally coupled Kuramoto system with $N=200$ oscillators as a function of the true coupling strength $K_{\mathrm{true}}$. (a) Ensemble distribution of the time-averaged order parameter $\langle R\rangle_t$. (b) Ensemble distribution of the fluctuation amplitude $\sigma_R$ evaluated over the same observation window.
    The box plots are based on 300 initial-phase realizations for each coupling strength, with fixed natural frequencies and no added observation noise. Boxes show the interquartile range (IQR), red lines indicate the medians, whiskers follow the $1.5$-IQR rule, and open circles denote outliers. The inset shows $\sigma_R$ on a logarithmic scale. The vertical dotted line marks the thermodynamic critical coupling $K_c \simeq 1.596$. The order-parameter fluctuations are enhanced near the synchronization transition and remain finite in the subcritical regime, indicating that the scalar trajectory $R(t)$ retains a nontrivial finite-size signal even below the synchronization threshold.
    }
    \label{fig:fluctuations}
\end{figure*}
\begin{figure*}[th]
    \centering
    \includegraphics[width=1\linewidth]{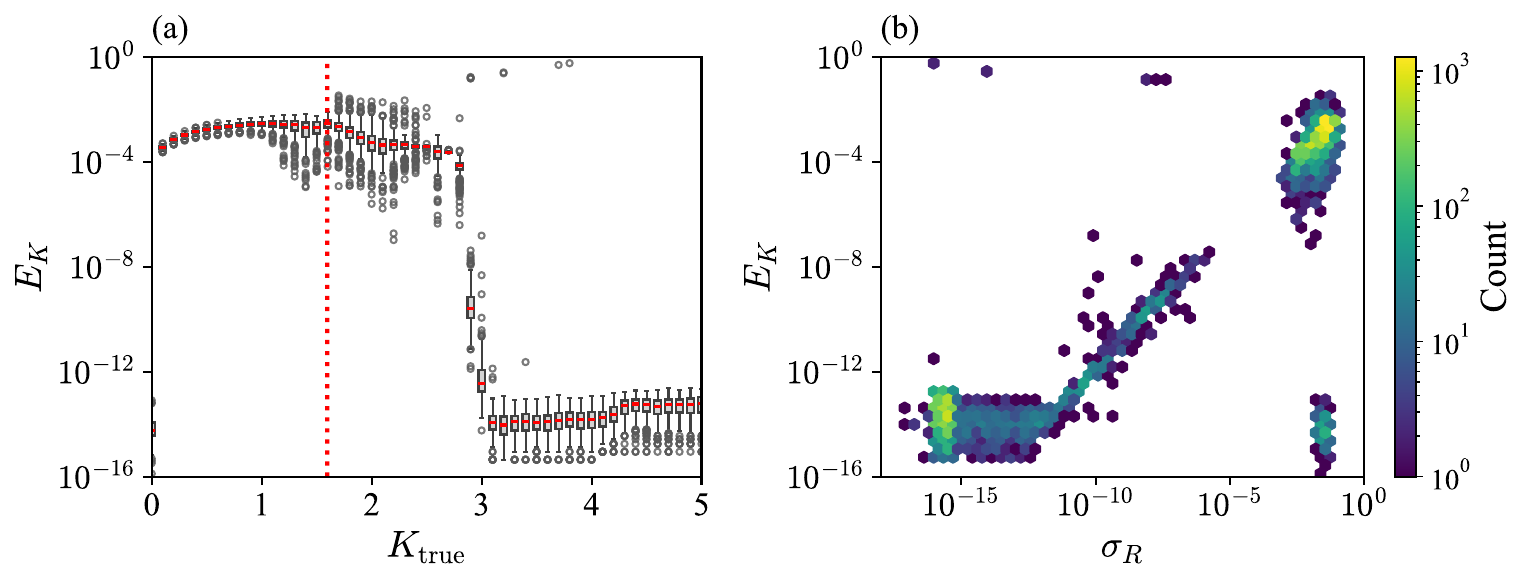}
    \caption{
    Ensemble statistics of the coupling-strength estimation for $N=200$.
    (a) Distribution of the residual estimation error,
    $E_K=|\hat{K}_{\rm f}-K_{\mathrm{true}}|$,
    as a function of the true coupling strength $K_{\mathrm{true}}$,
    where $\hat{K}_{\rm f}$ is defined as the temporal median of $\hat{K}(t)$
    over the final observation window.
    The error is shown on a logarithmic scale. Zero-error cases, if any, are excluded from the logarithmic visualization.
    Each box represents the interquartile range, the red line indicates the median,
    and open circles denote outliers.
    The vertical dotted line marks the critical coupling $K_c\simeq 1.596$.
    (b) Relation between the order-parameter fluctuation $\sigma_R$
    and the residual estimation error $E_K$.
    Both axes are shown on logarithmic scales.
    The color scale indicates the number of samples in each hexagonal bin,
    and white regions correspond to bins with no samples.
    Smaller fluctuations of $R(t)$ are generally associated with smaller residual estimation errors,
    whereas larger fluctuations lead to broader error distributions.
    }
    \label{fig:ensemble}
\end{figure*}
To evaluate the statistical performance of the estimator, we performed ensemble simulations over 300 realizations of the initial phases for each coupling strength.
For each realization, the final estimate was defined as the temporal median of $\hat{K}(t)$ over the final observation window $W$, $\hat{K}_{\rm f} = {\rm median}_{t\in W}\,\hat{K}(t)$.
The residual estimation error was then measured as $E_K = |\hat{K}_{\rm f}-K_{\rm true}|$.
The fluctuation amplitude of the order parameter was measured over the same window $W$ as
\begin{equation}
    \sigma_R
    =
    \sqrt{
    \left\langle R^2(t)\right\rangle_{t\in W}
    -
    \left\langle R(t)\right\rangle_{t\in W}^2
    }.
\end{equation}

To characterize the macroscopic signal available to the estimator, we first examine the finite-size dynamics of the order parameter itself.
Figure~\ref{fig:fluctuations} shows the ensemble distribution of the time-averaged order parameter $\langle R\rangle_t$ and its fluctuation amplitude $\sigma_R$ for a finite-size Kuramoto system with $N=200$ oscillators, as functions of the true coupling strength $K_{\mathrm{true}}$.
As expected, $\langle R\rangle_t$ increases across the synchronization transition near the thermodynamic critical coupling $K_c \simeq 1.596$.
For finite $N$, however, the order parameter does not vanish identically below $K_c$ but exhibits realization-dependent fluctuations.
These fluctuations become most pronounced near the transition point, where the susceptibility of the macroscopic state to finite-size effects is largest, and are strongly suppressed in the synchronized regime.
The presence of these finite-size fluctuations motivates the subsequent estimation analysis, where the EKF attempts to infer $K_{\mathrm{true}}$ from the scalar trajectory $R(t)$ even below the synchronization threshold.

\begin{figure*}[th]
    \centering
    \includegraphics[width=1\linewidth]{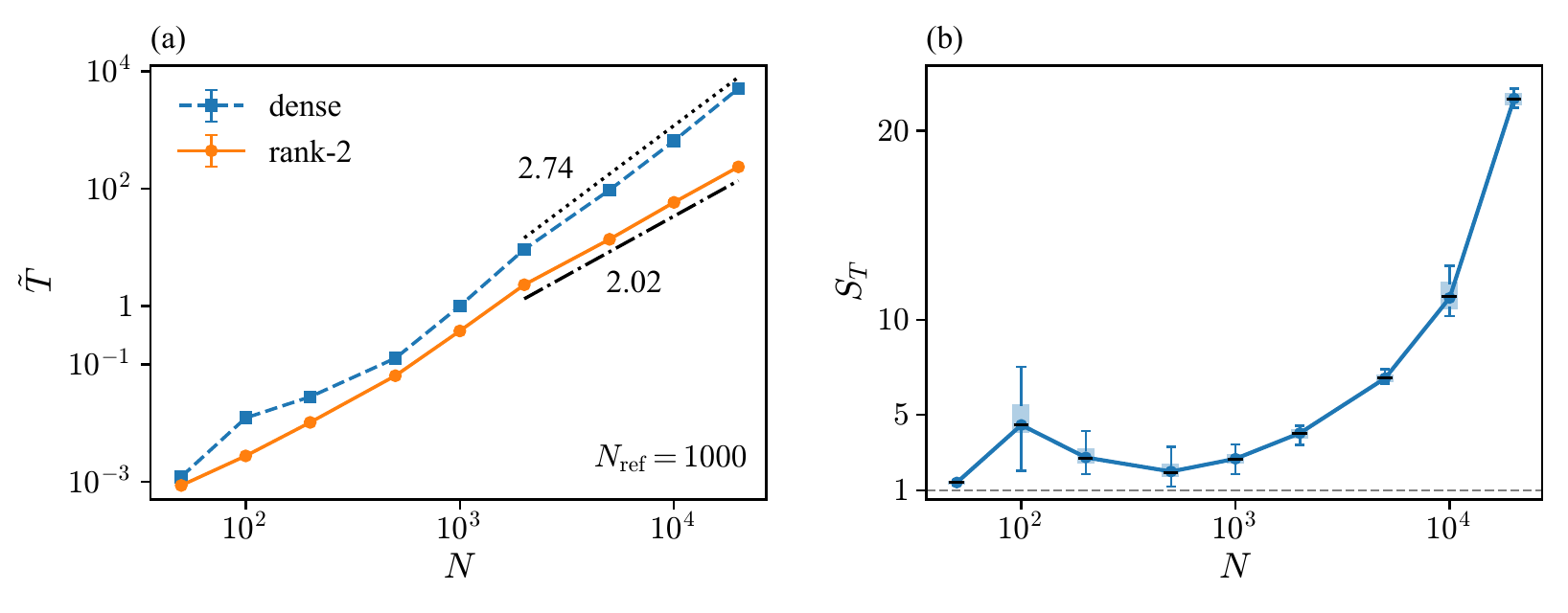}
    \caption{
    Computational scalability of the covariance-prediction step in the EKF. (a) Normalized computation time per covariance prediction, $\tilde{T}(N)=T_{\mathrm{step}}(N)/T_{\mathrm{dense}}(N_{\mathrm{ref}})$, for the dense covariance update and the proposed rank-2 update. Here, $N_{\mathrm{ref}}=10^3$. Markers indicate median runtimes, and error bars show the interquartile range. The dotted and dash-dotted lines show power-law fits over $2\times10^3\le N\le2\times10^4$, with exponents $2.74$ for the dense method and $2.02$ for the rank-2 method. (b) Computational speedup, $S_T(N)=T_{\mathrm{dense}}(N)/T_{\mathrm{rank\text{-}2}}(N)$. The dashed line indicates $S_T=1$. The nonmonotonic behavior at small $N$ reflects implementation-dependent overheads, whereas the increasing trend in the large-$N$ regime demonstrates the computational advantage of the rank-2 update.
    }
    \label{fig:efficiency}
\end{figure*}
Figure~\ref{fig:ensemble} summarizes the ensemble statistics.
The residual error remains bounded over the range of $K_{\rm true}$, and smaller fluctuations of $R(t)$ are generally associated with smaller residual estimation errors.
This relation suggests that finite-size order-parameter fluctuations, although small, still contain information about the underlying coupling strength when the initial microscopic state is known.


Finally, we tested the computational advantage of the rank-2 covariance prediction. For the computational benchmark, $N$ was varied to evaluate the scaling of the dense and rank-2 covariance prediction schemes. Figure~\ref{fig:efficiency} compares the normalized computation time per covariance prediction for the dense covariance update and the proposed rank-2 update. The normalized time is defined as $\tilde{T}(N) = {T_{\rm step}(N)}/{T_{\rm dense}(N_{\rm ref})}$, where $N_{\rm ref}=10^3$. We also measured the speedup ratio $S_T(N) = {T_{\rm dense}(N)}/{T_{\rm rank\text{-}2}(N)}$. A least-squares fit of $\log\tilde{T}$ versus $\log N$ over $2\times10^3\le N\le2\times10^4$ gives effective exponents of $2.74$ for the dense method and $2.02$ for the rank-2 method. Although the absolute clock time depends on the implementation and hardware, the increasing speedup ratio in the large-$N$ regime provides empirical support for the theoretical reduction of the covariance prediction cost from $\mathcal{O}(N^3)$ to $\mathcal{O}(N^2)$. The nonmonotonic behavior at small $N$ reflects implementation-dependent overheads.


\section{Discussion and Conclusion}
In this study, we demonstrated that an Extended Kalman Filter (EKF) framework can efficiently estimate the coupling strength of a finite-size Kuramoto network using the scalar macroscopic observation $R(t)$ solely, provided the initial microscopic phase configuration and natural frequencies are known. The estimator rapidly corrects an incorrect initial estimate of $K$ and remains bounded even in the subcritical regime, highlighting that finite-size fluctuations of the order parameter contain decodable information about the underlying coupling interactions. Furthermore, by rigorously deriving a rank-2 block covariance prediction of the Jacobian matrix, we reduced the computational complexity of covariance prediction from $\mathcal{O}(N^3)$ for the dense implementation to $\mathcal{O}(N^2)$. This provides a scalable foundation for tracking macroscopic-to-microscopic parameter inference in complex coupled systems.

The present framework assumes that the initial microscopic phase configuration and natural frequencies are known, which allows the filter to propagate a microscopic state estimate even though only the scalar order parameter is observed thereafter. With these priors, our numerical results demonstrate coupling-strength inference from unforced finite-size fluctuations even in the subcritical regime $K<K_c$. Experimental implementation would require measurement or preparation of the initial phases and calibration of individual natural frequencies from uncoupled time series, as illustrated by electrical-oscillator experiments~\cite{English2015}.

The present study assumes noiseless macroscopic observations in order to isolate the estimability of $K_{\rm true}$ from finite-size fluctuations alone. Extending the estimator to noisy observations, leveraging the general EKF formulation already presented in Appendix A, together with partially known initial states, uncertain natural frequencies, and more general network topologies, remains an important direction for future work. We note that the computational advantage reported here is established relative to a dense-matrix implementation of the same EKF, rather than through a direct comparison with the computational cost of Refs.~\cite{Kato2025,Yamaguchi2024,SmithGottwald2025,Choi2026}. Similarly, a matched estimation-accuracy comparison with these methods has not been performed, since our reported accuracy pertains to noiseless observations with exact microscopic priors.
 

\begin{acknowledgments}
This work was supported by Korea Research Institute for defense Technology planning and advancement (KRIT) - Grant funded by Defense Acquisition Program Administration (DAPA), South Korea (KRIT-CT-23-026, Integrated Underwater Surveillance Research Center for Adapting Future Technologies, 2023–2029).
\end{acknowledgments}

\appendix

\section{EKF Prediction and Measurement Update}
\label{app:ekf}

The true Kuramoto dynamics used to generate the observation time series $z_k=R(t_k)$
are integrated using a fourth-order Runge--Kutta method in the numerical simulations.
The Euler-discretized state-space model introduced in the main text is used by the EKF as its internal prediction model. Let the nonlinear process and observation models be
\begin{equation}
    \mathbf{x}_k = f(\mathbf{x}_{k-1})+\mathbf{w}_{k-1}, \qquad
    z_k = h(\mathbf{x}_k)+\epsilon_k,
\end{equation}
where $\mathbf{x}_k=[\theta_{1,k},\ldots,\theta_{N,k},K_k]^T$ is the augmented state vector,
$\mathbf{w}_{k-1}$ is the process noise, and $\epsilon_k\sim\mathcal{N}(0,R_c)$ is the observation noise.
The observation function is given by
\begin{equation}
    h(\mathbf{x}_k)=R_k=\sqrt{X_k^2+Y_k^2},
\end{equation}
where $X_k=\sum_{j=1}^{N}\cos\theta_{j,k}/N$, $Y_k=\sum_{j=1}^{N}\sin\theta_{j,k}/N$.

The EKF prediction step~~\cite{Kalman1960,Jazwinski1970} for state and covariance $P_k$ is
\begin{equation}
    \hat{\mathbf{x}}_{k|k-1}
    =
    f(\hat{\mathbf{x}}_{k-1|k-1}), \qquad
    P_{k|k-1}
    =
    F_{k-1}P_{k-1|k-1}F_{k-1}^{T}+Q_k,
\end{equation}
where
\begin{equation}
    F_{k-1}
    =
    \left.
    \frac{\partial f}{\partial \mathbf{x}}
    \right|_{\hat{\mathbf{x}}_{k-1|k-1}}
\end{equation}
is the Jacobian matrix of the process model (state transition model) evaluated at the posterior estimate from the previous step. $Q_k$ relates to the process noise in EKF model~~\cite{Kalman1960,Jazwinski1970}.

The measurement update is performed using $\nu_k = z_k-h(\hat{\mathbf{x}}_{k|k-1})$.
The Kalman gain is
\begin{equation}
    \mathcal{G}_k
    =
    P_{k|k-1}H_k^T
    \left(
    H_kP_{k|k-1}H_k^T+R_c
    \right)^{-1},
\end{equation}
where
\begin{equation}
    H_k
    =
    \left.
    \frac{\partial h}{\partial \mathbf{x}}
    \right|_{\hat{\mathbf{x}}_{k|k-1}}
\end{equation}
is the observation Jacobian. Since the observation depends only on the phases and not directly on $K_k$,
\begin{equation}
    H_k
    =
    \begin{bmatrix}
    h_{1,k} & h_{2,k} & \cdots & h_{N,k} & 0
    \end{bmatrix},
\end{equation}
with
\begin{equation}
    h_{i,k}
    =
    \frac{\partial R_k}{\partial \theta_{i,k}}
    =
    \frac{
    -X_k\sin\theta_{i,k}
    +
    Y_k\cos\theta_{i,k}
    }{
    N R_k
    }.
    \label{eq:obs_jacobian_element}
\end{equation}
In the nearly incoherent regime, $R_k$ can become very small. Therefore, in numerical implementation, a small lower bound is imposed on $R_k$ when evaluating Eq.~\eqref{eq:obs_jacobian_element}.

The posterior state estimate is updated as $\hat{\mathbf{x}}_{k|k} = \hat{\mathbf{x}}_{k|k-1} + \mathcal{G}_k\nu_k$.
The covariance update can be written in the standard form,
\begin{equation}
    P_{k|k}
    =
    (I-\mathcal{G}_kH_k)P_{k|k-1}.
\end{equation}
For numerical stability, the Joseph form may also be used:
\begin{equation}
    P_{k|k}
    =
    (I-\mathcal{G}_kH_k)
    P_{k|k-1}
    (I-\mathcal{G}_kH_k)^T
    +
    \mathcal{G}_kR_c\mathcal{G}_k^T.
\end{equation}

\section{Jacobian and Rank-2 Covariance Prediction}
\label{app:rank2}

The Jacobian matrix of the augmented state vector is partitioned as
\begin{equation}
    F_k
    =
    \begin{bmatrix}
    A_k & \mathbf{b}_k \\
    \mathbf{0}^T & 1
    \end{bmatrix},
\end{equation}
where $A_k\in\mathbb{R}^{N\times N}$ is the phase-interaction block and
$\mathbf{b}_k\in\mathbb{R}^{N}$ is the sensitivity of the phase update with respect to $K_k$.
For the Euler-discretized prediction model,
\begin{equation}
    \theta_{i,k+1}
    =
    \theta_{i,k}
    +
    \Delta t
    \left[
    \omega_i
    +
    K_k
    \left(
    Y_k\cos\theta_{i,k}
    -
    X_k\sin\theta_{i,k}
    \right)
    \right],
\end{equation}
the elements of $A_k$ are
\begin{equation}
    A_{ij}
    =
    \delta_{ij}
    \left[
    1
    -
    \Delta t K_k
    \left(
    X_k\cos\theta_{i,k}
    +
    Y_k\sin\theta_{i,k}
    \right)
    \right]
    +
    \frac{\Delta t K_k}{N}
    \left(
    \cos\theta_{i,k}\cos\theta_{j,k}
    +
    \sin\theta_{i,k}\sin\theta_{j,k}
    \right),
    \label{eq:A_element}
\end{equation}
and the sensitivity vector is
\begin{equation}
    b_i
    =
    \frac{\partial \theta_{i,k+1}}{\partial K_k}
    =
    \Delta t
    \left(
    Y_k\cos\theta_{i,k}
    -
    X_k\sin\theta_{i,k}
    \right).
    \label{eq:b_element}
\end{equation}

By defining
\begin{equation}
    \mathbf{c}
    =
    \begin{bmatrix}
    \cos\theta_{1,k} & \cdots & \cos\theta_{N,k}
    \end{bmatrix}^{T},
    \qquad
    \mathbf{s}
    =
    \begin{bmatrix}
    \sin\theta_{1,k} & \cdots & \sin\theta_{N,k}
    \end{bmatrix}^{T},
\end{equation}
the phase-interaction block can be decomposed exactly as
\begin{equation}
    A_k
    =
    D_k
    +
    \alpha
    \left(
    \mathbf{c}\mathbf{c}^{T}
    +
    \mathbf{s}\mathbf{s}^{T}
    \right),
    \qquad
    \alpha=\frac{\Delta t K_k}{N},
    \label{eq:rank2_decomposition}
\end{equation}
where $D_k$ is diagonal with elements
\begin{equation}
(D_k)_{ii}=1-\Delta t K_k\left(X_k\cos\theta_{i,k}+Y_k\sin\theta_{i,k}\right).
\label{eq:D_diagonal}
\end{equation}
Thus, the dense part of $A_k$ is not arbitrary but is a rank-2 perturbation of a diagonal matrix.

For brevity, we write $P_k=P_{k|k}$ and $P_k^{-}=P_{k|k-1}$.
To exploit this structure in the covariance prediction, we partition the covariance matrix as
\begin{equation}
    P_k
    =
    \begin{bmatrix}
    P_{\theta\theta,k} & \mathbf{p}_{\theta K,k} \\
    \mathbf{p}_{\theta K,k}^{T} & P_{KK,k}
    \end{bmatrix}.
\end{equation}
The prior covariance blocks are then updated as
\begin{align}
    P_{\theta\theta,k}^{-}
    &=
    A_{k-1}P_{\theta\theta,k-1}A_{k-1}^{T}
    +
    A_{k-1}\mathbf{p}_{\theta K,k-1}\mathbf{b}_{k-1}^{T}
    +
    \mathbf{b}_{k-1}\mathbf{p}_{\theta K,k-1}^{T}A_{k-1}^{T}
    +
    P_{KK,k-1}\mathbf{b}_{k-1}\mathbf{b}_{k-1}^{T}
    +
    Q_{\theta\theta},
    \label{eq:Ptheta_update} \\
    \mathbf{p}_{\theta K,k}^{-}
    &=
    A_{k-1}\mathbf{p}_{\theta K,k-1}
    +
    P_{KK,k-1}\mathbf{b}_{k-1},
    \label{eq:pK_update} \\
    P_{KK,k}^{-}
    &=
    P_{KK,k-1}+Q_K.
    \label{eq:PKK_update}
\end{align}

The dominant term in Eq.~\eqref{eq:Ptheta_update} is $A_{k-1}P_{\theta\theta,k-1}A_{k-1}^{T}$.
Let $P=P_{\theta\theta,k-1}$ and omit the time index for clarity. Using
$$
    A=D+\alpha(\mathbf{c}\mathbf{c}^{T}+\mathbf{s}\mathbf{s}^{T}),
$$
we obtain
\begin{equation}
    APA^T
    =
    DPD
    +
    \alpha
    \sum_{\mathbf{u}\in\{\mathbf{c},\mathbf{s}\}}
    \left[
    DP\mathbf{u}\,\mathbf{u}^{T}
    +
    \mathbf{u}\,\mathbf{u}^{T}PD
    \right] 
    +
    \alpha^2
    \sum_{\mathbf{u},\mathbf{v}\in\{\mathbf{c},\mathbf{s}\}}
    \mathbf{u}
    \left(
    \mathbf{u}^{T}P\mathbf{v}
    \right)
    \mathbf{v}^{T}.
    \label{eq:APA_rank2_expansion}
\end{equation}
The first term $DPD$ is evaluated by elementwise multiplication with the diagonal entries of $D$.
The remaining terms require only matrix-vector products such as $P\mathbf{c}$ and $P\mathbf{s}$,
scalar contractions such as $\mathbf{c}^{T}P\mathbf{s}$, and vector outer products.
Therefore, no dense matrix-matrix multiplication is required, and the covariance prediction step is evaluated with leading computational cost $\mathcal{O}(N^2)$ instead of $\mathcal{O}(N^3)$.

\section{EKF Numerical Implementation Details}
\label{app:impl}
The initial covariance matrix was set to
$$
    P_{\theta\theta,0}=10^{-6}I,
    \qquad
    P_{KK,0}=10^{-2},
$$
with zero initial cross-covariance between the phase variables and $K$.
The process noise covariance was taken to be zero for all phase variables and nonzero only for the coupling-strength component,
$$
    Q_{\theta\theta}=\mathbf{0},
    \qquad
    Q_K=10^{-10}.
$$
This corresponds to modeling $K$ as a weak random-walk parameter while treating the phase evolution as deterministic within the EKF prediction model.
Unless otherwise stated, the measurement noise variance was set to $R_c=0$, so that the scalar observation $z_k$ corresponds directly to the synthetic order parameter $R(t_k)$.
After each measurement update, the estimated parameter $\hat{K}$ was projected onto the interval $[0,20]$.

\section{Microscopic Phase-Tracking Visualization}
\label{app:phase}
To visualize the microscopic phase-tracking performance, we compare the true and estimated phases of individual oscillators.
Since the oscillator phase is periodic, the phase difference is evaluated modulo $2\pi$ as
\begin{equation}
\Delta\theta_i(t)
=
\mathrm{arg}
\left[
e^{i(\theta_i(t)-\hat{\theta}_i(t))}
\right],
\end{equation}
where $\mathrm{arg}(\cdot)\in(-\pi,\pi]$.
Figure~\ref{fig:phase_tracking} shows $\Delta\theta_i(t)$ for individual oscillators in a representative case with $K_{\mathrm{true}}=1$ and $N=200$.
The dashed horizontal lines indicate $\Delta\theta_i=\pm0.01$ rad.
Most phase deviations remain within this range over the simulated interval, corresponding to less than $1\%$ of the full phase range $2\pi$.
\setcounter{figure}{0}
\renewcommand{\thefigure}{D\arabic{figure}}
\begin{figure}[t!]
    \centering
    \includegraphics[width=0.7\linewidth]{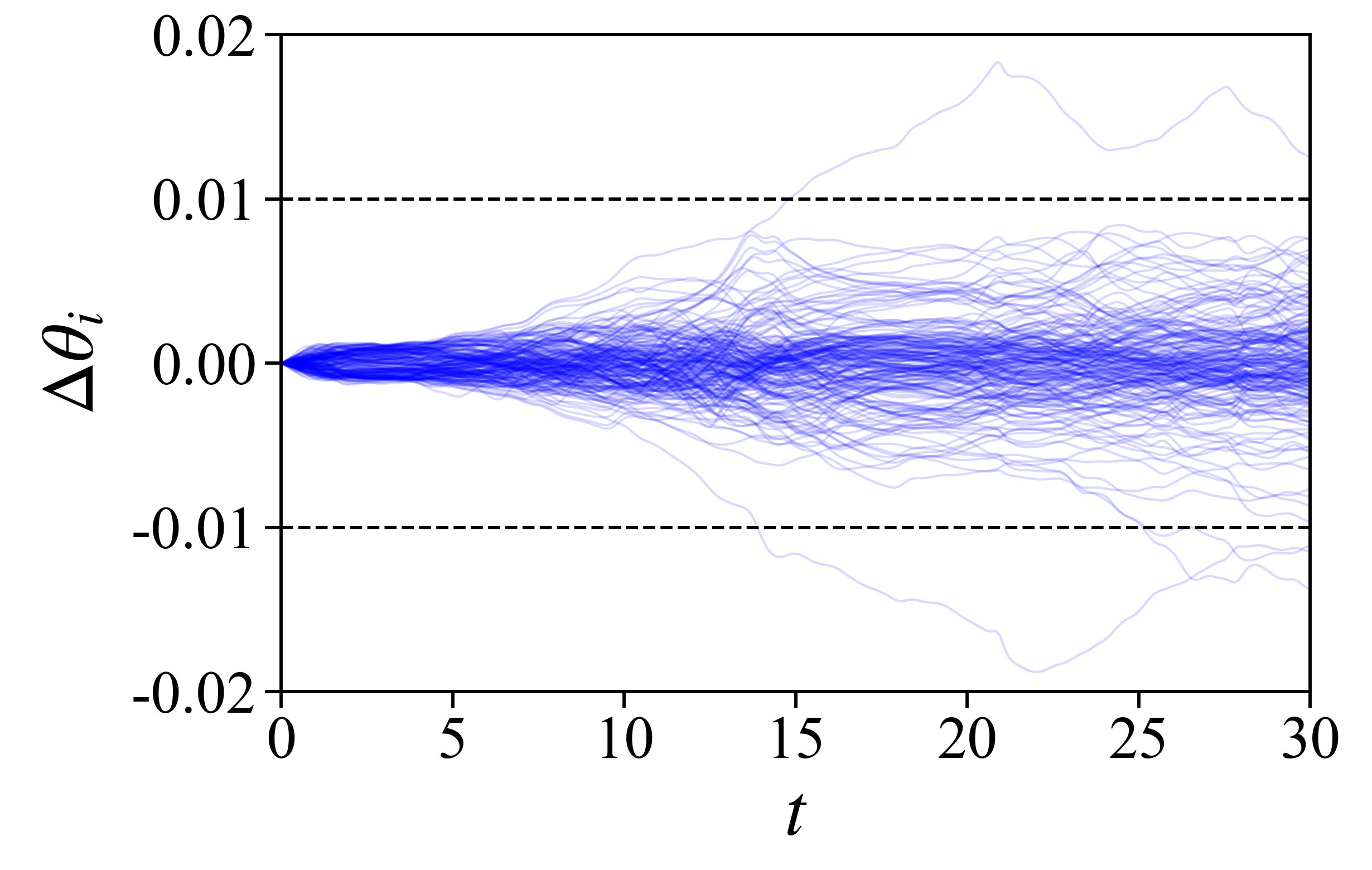}
    \caption{
    Microscopic phase-tracking visualization of the EKF for $K_{\mathrm{true}}=1$ with $N=200$.
    Each curve represents the phase difference $\Delta\theta_i(t)$ between the true and estimated phase of an individual oscillator.
    The dashed horizontal lines indicate $\Delta\theta_i=\pm0.01$ rad.
    Although only the scalar order parameter $R(t)$ is used for correction after initialization, most phase deviations remain within $|\Delta\theta_i|\lesssim0.01$ rad over the simulated interval, corresponding to less than $1\%$ of the full phase range $2\pi$.
    }
    \label{fig:phase_tracking}
\end{figure}

\FloatBarrier

\end{document}